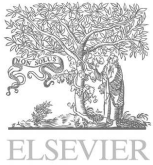

# The Microcalorimeter Arrays for a Rhenium Experiment (MARE): a next-generation calorimetric neutrino mass experiment

A. Monfardini,[f] C. Arnaboldi,[e] C. Brofferio,[e] S. Capelli,[e] F. Capozzi,[e] O. Cremonesi,[e] C. Enss,[c] E. Fiorini,[e,*] A. Fleischmann,[c] L. Foggetta,[d] G. Gallinaro,[a] L. Gastaldo,[c] F. Gatti,[a] A. Giuliani,[d] P. Gorla,[e] R. Kelley,[b] C.A. Kilbourne,[b] B. Margesin,[f] D. McCammon,[g] C. Nones,[e] A. Nucciotti,[e,†] M. Pavan,[e] M. Pedretti,[d] D. Pergolesi,[a] G. Pessina,[e] F.S. Porter,[b] M. Prest,[d] E. Previtali,[e] P. Repetto,[a] M. Ribeiro-Gomez,[a] S. Sangiorgio,[d] M. Sisti,[e]

[a]*INFN sez. Genova and Università di Genova, Dipartimento di Fisica, via Dodecaneso 33, I-16146 Genova, Italy*

[b]*NASA Goddard Space Flight Center, Greenbelt, MD 20771, USA*

[c]*Universität Heidelberg, Kirchhoff-Institut für Physik, INF 227, Heidelberg D-69120, Germany*

[d]*Università dell'Insubria, Dipartimento di Fisica e Matematica and INFN sez. Milano, via Valleggio 11, I-22100 Como, Italy*

[e]*INFN sez. Milano and Università di Milano-Bicocca, Dipartimento di Fisica, piazza Della Scienza 3, I-20126 Milano, Italy*

[f]*ITC-IRST and INFN gruppo di Trento, via Sommarive 18, I-38050 Povo (TN), Italy*

[g]*University of Wisconsin, Physics Department, 1150 University Ave. Madison, WI 53706, USA*

**Abstract**

Neutrino oscillation experiments have proved that neutrinos are massive particles, but can't determine their absolute mass scale. Therefore the neutrino mass is still an open question in elementary particle physics. An international collaboration is growing around the project of Microcalorimeter Arrays for a Rhenium Experiment (MARE) for directly measuring the neutrino mass with a sensitivity of about $0.2 eV/c^2$. Many groups are joining their experiences and technical expertise in a common effort towards this challenging experiment. We discuss the different scenarios and the impact of MARE as a complement of KATRIN. © 2001 Elsevier Science. All rights reserved

---

[*] Collaboration spokesman. Tel. +39 02 64482424. e-mail: ettore.fiorini@mib.infn.it

[†] Corresponding author. Tel. +39 02 64482428. e-mail: angelo.nucciotti@mib.infn.it

*Keywords*: neutrino mass, beta decay, thermal detectors, $^{187}$Re.

*PACS*: 23.40.Bw; 14.60.Pq

## 1. Introduction and Science case

As a result of a big effort of the neutrino physics community in the last decades, flavor oscillations in atmospheric, solar and reactor neutrinos have been observed and studied [1]. Neutrinos have a mass. This represents a clear evidence for new Physics beyond the Standard Model, and direct experiments sensitive to the mass absolute scale have been upgraded from "fishing" to "hunting".

The implications of an $m_\nu$ measurement would be terrific in Fundamental Physics (Grand Unified Theories) and Cosmology (large scale structure, dark matter). Even if debatable, a first hint came from the claimed observation of neutrinoless double beta decay in $^{76}$Ge. According to the estimated $T_{1/2}$, the range of effective Majorana mass is $m_{ee} = 0.1 \div 0.9$ eV ($3\sigma$), allowing a 50% uncertainty in the nuclear matrix [2]. Indirect cosmological indications, on the other hand, suggest $\Sigma m_\nu < 0.42$ eV/c$^2$ at 95%C.L.[3].

The new frontier in neutrino physics is thus to set up a really direct neutrino mass experiment (i.e. single $\beta$ decay) sensitive enough to explore the sub-eV/c$^2$ range. The best result so far is $m_\nu < 2.2$ eV/c$^2$ [4] obtained by studying the $^3$H beta end-point with an electrostatic spectrometer. KATRIN [5], the next generation spectrometer, is designed to reach 0.2eV/c$^2$ $m_\nu$ sensitivity[1]. The KATRIN 1$^{st}$ run should begin in the second half of 2009. KATRIN will probably approach the ultimate spectrometers limit.

An alternative technique able to eliminate the systematic effects related to the separation between the source and the detector is the calorimetric one. The absorber of a calorimeter acts both as a source and detector. A thermistor records the transient $\Delta T$ induced by the single $^{187}$Re nucleus decay. $^{187}$Re is the lowest-$Q$ known beta decaying nucleus, and by far the most interesting for neutrino mass searches.

---

[1] Since the measured $\Delta m^2$ are exceedingly small, $m_\nu$ represents in this context the $m_1 \approx m_2 \approx m_3$ scale in the quasi-degenerate pattern. If $m_1 \ll m_2 \ll m_3$ (hierarchical pattern), even the largest mass $m_3$ will remain below any proposed (direct) experiment sensitivity.

The best published calorimetric limit is $m_\nu < 15$ eV/c$^2$ obtained by the small MIBETA array [6]. A large array of at least $10^4$ next-generation thermal detectors is required to improve this limit by two orders of magnitudes. Starting from MIBETA and MANU [7] experiences, an international collaboration is growing on the Microcalorimeter Arrays for a Rhenium Experiment (MARE) project [8]. MARE is a two-stage effort. The first phase, coming to an end before the KATRIN first run, will result in an order of magnitude sensitivity improvement down to about 2eV/c$^2$. During the first phase data taking (2-3 years) an intense detector R&D activity is needed to prepare the 2$^{nd}$ phase and finally reach 0.2eV/c$^2$ sensitivity.

## 2. The MARE phase I: simulations and preliminary results

A Montecarlo (MC) code has been developed to explore different configurations for MARE phases I and II. The main parameters are: $N_{ev}$ total number of beta decays, $\Delta E$ detector energy resolution, $f_{pup}$ pile-up fraction related to the beta rate per detector $A_\beta$ and the pulse rise time $\tau_R$: $f_{pup} \approx A_\beta \cdot \tau_R$. A new feature is the introduction of solid-state effects (BEFS).

From MC, assuming $\Delta E = 10$ eV and $f_{pup} = 5 \cdot 10^{-5}$, we know that more than $5 \cdot 10^9$ beta decays are needed to achieve the scientific goals of MARE phase I.

Two parallel approaches will satisfy the requirements.

### 2.1. MIBETA2: semiconductor (SC) thermistors

The new MIBETA array (about 200 detectors) is based on semiconductor thermistors and dielectric (AgReO$_4$) absorbers. Three different thermistor technologies are being investigated to optimize the energy resolution and reduce the pile-up:

- ITC-IRST arrays of implanted, micromachined silicon thermistors. The thermistor ionic implantation is based on the single devices extensively tested in the past. A production run is ongoing, and the first cryogenics tests are expected in October 2005;

- NASA-GSFC Si micromachined arrays (XRS). Tests have already been performed in Milano with $AgReO_4$ absorbers. Despite the still non optimized electronic chain, this array seems already a good baseline for MARE phase I;

- NTD Germanium arrays. In theory they are preferred as far as the electro-thermal model is concerned. However, the scaling up to hundreds of detectors seems more difficult to achieve.

The technology providing the best results will be adopted from the SC side.

*2.2. MANU2 : Transition Edge Sensors (TES)*

MANU2 is a made by metallic rhenium single crystals cooled to 60mK. The thermistors are Ir-Au TES realized by laser ablation techniques on a silicon substrate. The energy resolution is about 10eV FWHM, with pulse rise time of 25-30μs.

A read-out electronics based on transformers and low-temperature JFETs is adopted due to the difficulties in acquiring commercial SQUIDs with the required noise performances. The detectors have been suitably matched to the new front-end. The number of detectors is around 150 for an expected sensitivity of about $1eV/c^2$ in 1 to 2 years measurement time.

**3. Phase II: perspectives and conclusions**

The main scientific goal will be achieved only if the phase II will actually follow. The number of events for phase II has to improve by another four orders of magnitudes to $N_{ev} \approx 10^{14}$. By extrapolating the present technology trend, we believe that this can't be easily obtained by merely increasing the number of detectors. The baseline array size for a single MARE phase II detector module is $10^4$, and several modules can be run in parallel. The single detector decay rate has then to be maximized without influencing the pile-up fraction. This means bigger absorbers and faster pulses. In addition, the required energy resolution is of the order of 1÷5eV. These performances are, even if not dramatically, beyond the capabilities of the existing, single devices. TES or MMC (Metallic Magnetic Calorimeters) equipped with multiplexed SQUID read-out seem the most promising detectors to bridge this gap.

The phase II data taking will begin in 2010, about one year after the end of the phase I. The two-steps approach ensures a preliminary, safe understanding of the systematic effects. In addition, the present spectrometers $m_\nu$ limits will be explored with a possibly safer technique before the KATRIN results.

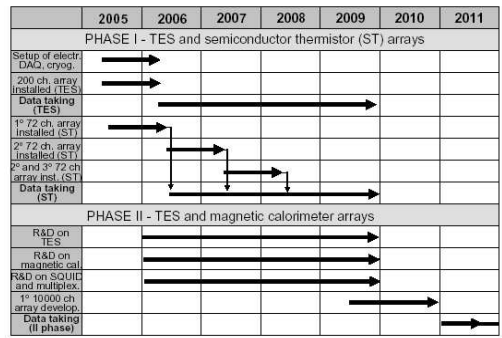

Figure 1. MARE phases I and II schedules.

In conclusion, a medium-to-large ($N_{DET} \approx 5 \cdot 10^4$) calorimetric experiment is required to explore the interesting sub-$eV/c^2$ absolute neutrino mass range. KATRIN, the next generation spectrometer, is probably going to achieve sub-$eV/c^2$ sensitivities a bit earlier, but also reaching the technique ultimate limit. Even in case of a $m_\nu$ positive detection, a quasi-simultaneous calorimetric cross-check will represent an important confirmation.

In case the neutrino mass pattern is hierarchical, futuristic developments in calorimeters technologies toward CCD-like arrays ($N > 10^6$) will be required to make a direct determination possible.